\documentstyle[preprint,eqsecnum,epsf,epsfig,aps,floats,tighten]{revtex} 
\input{epsf}       

\def\Missing#1#2{{\mbox{$#1\kern-0.57em\raise0.19ex\hbox{/}_{#2}$}}}

\def\vMissing#1#2{\ifmmode
            \vec{#1}\kern-0.57em\raise.19ex\hbox{/}_{#2}
         \else
            {{\mbox{$\vec{#1}\kern-0.57em\raise.19ex\hbox{/}_{#2}$}}}
         \fi}

\newcommand{\pbarp}{\mbox{$p\overline{p}$}}

\newcommand{\inb}{\mbox{nb$^{-1}$}}

\begin{document}
%
%
\title{
Multiple Jet Production at Low Transverse Energies in \pbarp \ Collisions
at $\sqrt{s}$ = 1.8 TeV}

%
\author{                                                                      
V.M.~Abazov,$^{22}$                                                           
B.~Abbott,$^{56}$                                                             
A.~Abdesselam,$^{11}$                                                         
M.~Abolins,$^{49}$                                                            
V.~Abramov,$^{25}$                                                            
B.S.~Acharya,$^{17}$                                                          
D.L.~Adams,$^{54}$                                                            
M.~Adams,$^{36}$                                                              
S.N.~Ahmed,$^{21}$                                                            
G.D.~Alexeev,$^{22}$
% ...
\\                                                                            
\vskip 0.30cm                                                                 
\centerline{(D\O\ Collaboration)}                                             
\vskip 0.30cm                                                                 
}                                                                             
\address{                                                                     
\centerline{$^{1}$Universidad de Buenos Aires, Buenos Aires, Argentina}       
\centerline{$^{2}$LAFEX, Centro Brasileiro de Pesquisas F{\'\i}sicas,         
                  Rio de Janeiro, Brazil}                                     
\centerline{$^{3}$Universidade do Estado do Rio de Janeiro,                   
                  Rio de Janeiro, Brazil}                                     
\centerline{$^{4}$Institute of High Energy Physics, Beijing,                  
                  People's Republic of China}                                 
\centerline{$^{5}$Universidad de los Andes, Bogot\'{a}, Colombia}             
\centerline{$^{6}$Charles University, Center for Particle Physics,            
                  Prague, Czech Republic}                                     
\centerline{$^{7}$Institute of Physics, Academy of Sciences, Center           
                  for Particle Physics, Prague, Czech Republic}               
\centerline{$^{8}$Universidad San Francisco de Quito, Quito, Ecuador}         
\centerline{$^{9}$Institut des Sciences Nucl\'eaires, IN2P3-CNRS,             
                  Universite de Grenoble 1, Grenoble, France}                 
\centerline{$^{10}$CPPM, IN2P3-CNRS, Universit\'e de la M\'editerran\'ee,     
                  Marseille, France}                                          
\centerline{$^{11}$Laboratoire de l'Acc\'el\'erateur Lin\'eaire,              
                  IN2P3-CNRS, Orsay, France}                                  
\centerline{$^{12}$LPNHE, Universit\'es Paris VI and VII, IN2P3-CNRS,         
                  Paris, France}                                              
\centerline{$^{13}$DAPNIA/Service de Physique des Particules, CEA, Saclay,    
                  France}                                                     
\centerline{$^{14}$Universit{\"a}t Mainz, Institut f{\"u}r Physik,            
                  Mainz, Germany}                                             
\centerline{$^{15}$Panjab University, Chandigarh, India}                      
\centerline{$^{16}$Delhi University, Delhi, India}                            
\centerline{$^{17}$Tata Institute of Fundamental Research, Mumbai, India}     
\centerline{$^{18}$Seoul National University, Seoul, Korea}                   
\centerline{$^{19}$CINVESTAV, Mexico City, Mexico}                            
\centerline{$^{20}$FOM-Institute NIKHEF and University of                     
                  Amsterdam/NIKHEF, Amsterdam, The Netherlands}               
\centerline{$^{21}$University of Nijmegen/NIKHEF, Nijmegen, The               
                  Netherlands}                                                
\centerline{$^{22}$Joint Institute for Nuclear Research, Dubna, Russia}       
\centerline{$^{23}$Institute for Theoretical and Experimental Physics,        
                   Moscow, Russia}                                            
\centerline{$^{24}$Moscow State University, Moscow, Russia}                   
\centerline{$^{25}$Institute for High Energy Physics, Protvino, Russia}       
\centerline{$^{26}$Lancaster University, Lancaster, United Kingdom}           
\centerline{$^{27}$Imperial College, London, United Kingdom}                  
\centerline{$^{28}$University of Arizona, Tucson, Arizona 85721}              
\centerline{$^{29}$Lawrence Berkeley National Laboratory and University of    
                  California, Berkeley, California 94720}                     
\centerline{$^{30}$University of California, Davis, California 95616}         
\centerline{$^{31}$California State University, Fresno, California 93740}     
\centerline{$^{32}$University of California, Irvine, California 92697}        
\centerline{$^{33}$University of California, Riverside, California 92521}     
\centerline{$^{34}$Florida State University, Tallahassee, Florida 32306}      
\centerline{$^{35}$Fermi National Accelerator Laboratory, Batavia,            
                   Illinois 60510}                                            
\centerline{$^{36}$University of Illinois at Chicago, Chicago,                
                   Illinois 60607}                                            
\centerline{$^{37}$Northern Illinois University, DeKalb, Illinois 60115}      
\centerline{$^{38}$Northwestern University, Evanston, Illinois 60208}         
\centerline{$^{39}$Indiana University, Bloomington, Indiana 47405}            
\centerline{$^{40}$University of Notre Dame, Notre Dame, Indiana 46556}       
\centerline{$^{41}$Iowa State University, Ames, Iowa 50011}                   
\centerline{$^{42}$University of Kansas, Lawrence, Kansas 66045}              
\centerline{$^{43}$Kansas State University, Manhattan, Kansas 66506}          
\centerline{$^{44}$Louisiana Tech University, Ruston, Louisiana 71272}        
\centerline{$^{45}$University of Maryland, College Park, Maryland 20742}      
\centerline{$^{46}$Boston University, Boston, Massachusetts 02215}            
\centerline{$^{47}$Northeastern University, Boston, Massachusetts 02115}      
\centerline{$^{48}$University of Michigan, Ann Arbor, Michigan 48109}         
\centerline{$^{49}$Michigan State University, East Lansing, Michigan 48824}   
\centerline{$^{50}$University of Nebraska, Lincoln, Nebraska 68588}           
\centerline{$^{51}$Columbia University, New York, New York 10027}             
\centerline{$^{52}$University of Rochester, Rochester, New York 14627}        
\centerline{$^{53}$State University of New York, Stony Brook,                 
                   New York 11794}                                            
\centerline{$^{54}$Brookhaven National Laboratory, Upton, New York 11973}     
\centerline{$^{55}$Langston University, Langston, Oklahoma 73050}             
\centerline{$^{56}$University of Oklahoma, Norman, Oklahoma 73019}            
\centerline{$^{57}$Brown University, Providence, Rhode Island 02912}          
\centerline{$^{58}$University of Texas, Arlington, Texas 76019}               
\centerline{$^{59}$Texas A\&M University, College Station, Texas 77843}       
\centerline{$^{60}$Rice University, Houston, Texas 77005}                     
\centerline{$^{61}$University of Virginia, Charlottesville, Virginia 22901}   
\centerline{$^{62}$University of Washington, Seattle, Washington 98195}       
\centerline{$^{63}$Petersburg Nuclear Physics Institute, Gatchina, Russia}
}                                                                             

%
%
%
%

\maketitle

%
%
\begin{abstract}
We present data on multiple production of jets with transverse energies
near 20 GeV in {\pbarp} collisions at $\sqrt{s}$ = 1.8 TeV.  QCD
calculations in the parton-shower approximation of {\sc pythia}
and {\sc herwig} and the next--to--leading order approximation of
{\sc jetrad} are compared to the data for one, two, three, and
four jet inclusive production.  Transverse energy spectra and multiple 
jet angular and summed transverse-energy distributions are adequately
described by the shower approximation while next--to--leading
order calculations describe the data poorly.
\end{abstract}


\newpage

\normalsize

\vfill\eject

\section {Introduction}
\label{sec:intro}

The study of multiple jet production at high transverse energy was
a goal of the 1993--1995 run of the Fermilab Tevatron collider,
and the results have been compared with leading-order QCD
predictions by both the CDF\cite{ref_1} and D\O \cite{ref_2}
collaborations. These high-$E_T$ data, where $E_T$ is the
transverse energy of the jet, are described satisfactorily by
complete tree-level leading order $2\rightarrow N$ QCD
calculations\cite{ref_3} and by the {\sc herwig} parton-shower
Monte Carlo\cite{ref_4} program. This kinematic region is described by
$Q^2/\hat s \approx 1 $, where $Q^2$ is the square of the momentum
transfer between partons (which we set equal to $E_{T}^2$), and
$\hat s$ is the square of the partonic center of mass energy. 
In this paper, we describe jet production measurements at significantly 
lower values of $E_T$ where detailed measurement of jet production in 
this kinematic region can provide information on the evolution of 
higher-order jet processes. In the same low $E_T$ region the D\O \ 
collaboration has previously reported the ratio of the inclusive 
three-jet to the inclusive two-jet cross section as a function of the 
scalar sum of jet transverse energies ($H_T=\sum E_T $) with $E_T > $ 
20~GeV\cite{ref_5}. The ratio data can be described by the {\sc jetrad} 
next-to-leading order Monte Carlo\cite{ref_13} program. In this paper 
we make comparisons between Monte Carlo and several characteristics of 
multiple jet events including the leading jet transverse energy, the 
relative azimuthal angle between jets, and the summed vector 
transverse momenta of jets.

\section {Data Sample and Corrections}
\label{sec:Data}

The data were collected with the
D\O \ detector at a proton-antiproton center-of-mass energy $\sqrt{s}$
= 1.8 TeV. Jets were identified using the liquid-argon uranium
calorimeters, which have segmentation of $\Delta \eta \times
\Delta \phi =0.1\times 0.1$, where pseudorapidity 
$\eta = -\ln \tan {{\theta}\over{2}}$, 
$\theta$ is the polar angle, and $\phi $ is
azimuthal angle\cite{ref_7}. At least one calorimeter trigger
tower ($\Delta \eta \times \Delta \phi =0.2\times 0.2$) with
$E_T \ge$ 2 GeV was required by a hardware trigger, and at
least one jet with $E_T \ge $ 12 GeV was required by a subsequent 
software trigger\cite{ref_11_0}. Jets were reconstructed using a
fixed cone algorithm with radius $\Delta {\cal {R}}=\sqrt{\Delta
\eta ^2+\Delta \phi ^2}=0.7$ in $\eta -\phi $
space\cite{ref_11_0}. The jet reconstruction threshold was
$E_T=8$ GeV. If two jets overlapped and the shared
transverse energy was more than 50\% of the transverse energy of
the lower-$E_T$ jet, the jets were merged; otherwise they
were split into two jets. The integrated luminosity of this data
sample is 2.0$\pm $0.3 {\inb}. Instantaneous luminosity was
restricted to be below 3 $\times 10^{30}~$cm$^{-2}$s$^{-1}$ to minimize
the number of multiple {\pbarp} interactions in a single beam crossing.

To provide events of high quality, online and offline selection
criteria suppressed multiple interactions, the cosmic ray background,
and spurious jets\cite{ref_11_0}. Jets were restricted to the
pseudorapidity interval $|\eta | \le 3$. The primary vertex of
each event (reconstructed from time-of-flight as measured by
scintillation counters\cite{ref_7}) was required to be within 50 cm 
of the detector center.

Jet energies have been corrected for calorimeter response, shower
development, and \newline various sources of noise\cite{ref_8}.
These corrections constitute the largest source of systematic
uncertainty on the jet cross section. Typical values of the jet
energy correction are (15--30)\%, with an uncertainty of (2--4)\%. 
In our study, we consider jets with $E_T > $ 20~GeV; for an inclusive
$n$-jet event, the $n$ jets with the maximum $E_T$ (the leading jets) 
must have transverse energy above the threshold value. For example, 
a 3-jet event must have at least 3 jets above 20 GeV. The trigger 
efficiency is 0.85 for the inclusive ($n$ = 1) jet sample for energies 
near threshold, rising rapidly to unity at larger $E_T$. The efficiency
is essentially unity for $n > 1$.

To compare with data, Monte Carlo (MC) events were generated using
the {\sc pythia} 6.127\cite{ref_pythia}, {\sc herwig}
5.9\cite{ref_4}, and {\sc jetrad}\cite{ref_13} programs. {\sc
pythia} and {\sc herwig} simulate particle-level jets in the
parton-shower approximation. {\sc jetrad} simulates jets in the
next-to-leading order approximation. To simulate detector
resolution effects, the MC jet transverse energies were smeared
with the experimentally determined jet energy
resolution\cite{ref_8}, which is $\approx $ 20\% at $E_T$ = 20
GeV. Jet angular smearing used $\eta $ and $\phi $ resolutions
obtained by a MC simulation of the calorimeter response using {\sc
herwig} 5.9 and {\sc geant} \cite{ref_geant}. These resolutions
are $\approx 0.08$ at $E_T$ = 20 GeV. In {\sc pythia} and {\sc
herwig}, jets were reconstructed at the particle level using the
D\O\ algorithm, and in {\sc jetrad}, at the parton level, using
the Snowmass algorithm\cite{ref_11_1}.

\section {Leading jet $E_T$ distributions and systematic
uncertainties}
\label{sec:Lead}

\begin{figure}[h]
\vspace {-4.0cm}
\begin{center}
\mbox{\hspace*{0.0cm}\epsfig{figure=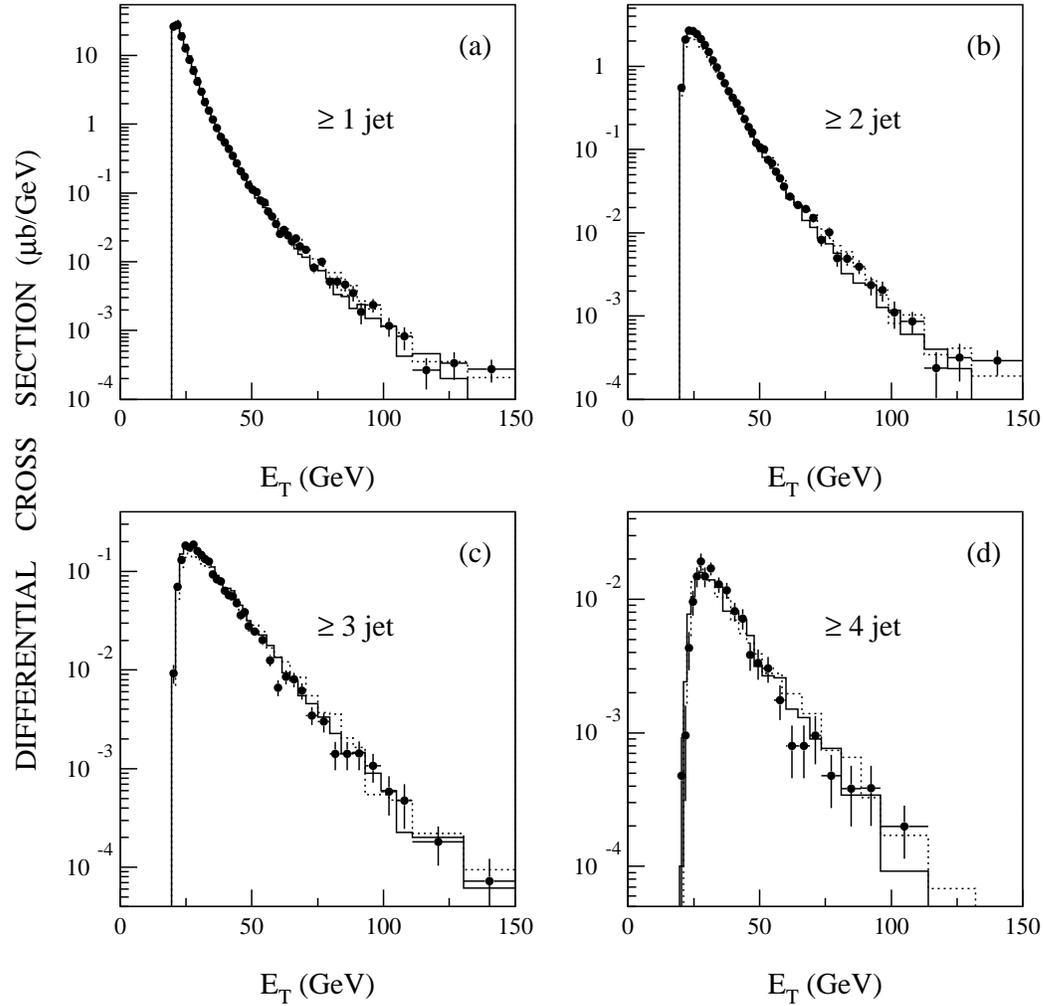,width=15cm}}
\vspace {0.5cm}
\caption{
The transverse energy distributions of the leading
jet for (a) single-inclusive, (b) two-jet inclusive, (c) three-jet
inclusive, and (d) four-jet inclusive events. Solid histograms show the
{\sc pythia} simulation normalized (with a factor of 0.75)
to the inclusive two-jet sample for $E_T$ $>$ 40 GeV. Dotted histograms
are similarly normalized {\sc herwig} results (increased by a factor of 1.6).
}
\label{et_angsm}
\end{center}
\end{figure}

Distributions in transverse energy for the leading jet for inclusive 
$n$=1 to $n$=4 jet events are shown in Fig.~\ref{et_angsm},
along with the results from {\sc pythia} and {\sc herwig}
simulations. In these and all other plots, the data have been
corrected for inefficiencies and energy calibration, but not for
contributions from the underlying event. All simulated
distributions have been smeared with energy and angular
resolutions. Also to describe the data quantitatively, we normalize 
the theory (with a factor of 0.75 for {\sc pythia} and 1.6 for 
{\sc herwig}) to the observed two-jet inclusive cross section in 
Fig.~\ref{et_angsm}(b) for $E_T > 40 \rm ~GeV$.

\begin{figure}[h]
\vspace {-4.0cm}
\begin{center}
\mbox{\hspace*{0.0cm}\epsfig{figure=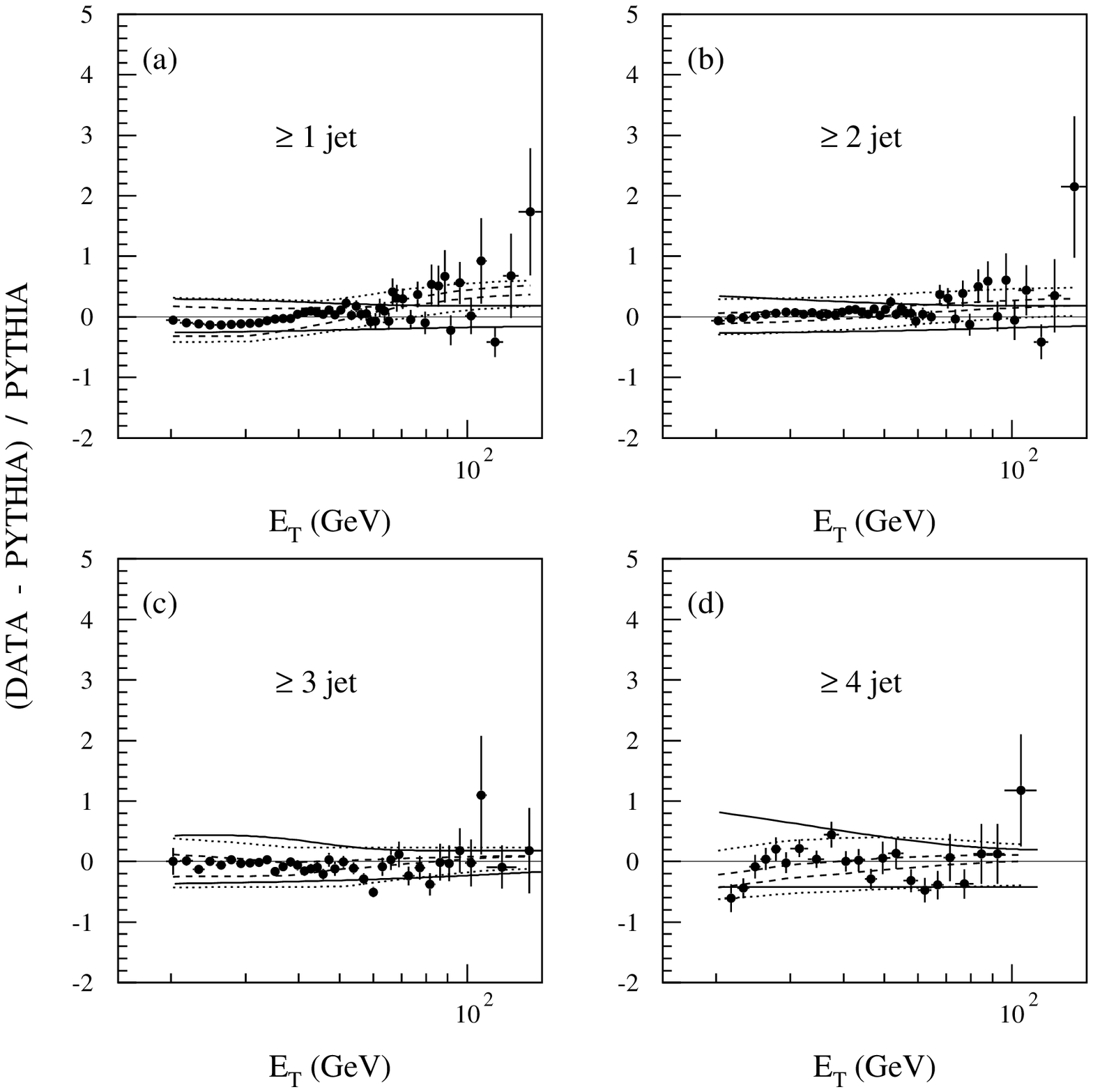
,width=15cm}}
\vspace {0.5cm}
\caption{
(Data $-$ {\sc pythia})/{\sc pythia} as a function of
the transverse energy of the leading jet for (a) single-jet inclusive,
(b) two-jet inclusive, (c) three-jet inclusive, and (d) four-jet
inclusive event samples. The relative systematic uncertainties in the 
cross section corresponding to the energy calibration added in 
quadrature with 15\% uncertainty in luminosity are shown by the solid 
lines. The uncertainty in the ratio (Data $-$ MC) / MC from energy and 
angle smearing is shown by the dashed lines. The total uncertainty on the 
ratio is shown by the dotted lines.
}
\label{et_prim_sys_angsm}
\end{center}
\end{figure}

\begin{figure}[h]
\vspace {-4.0cm}
\begin{center}
\mbox{\hspace*{0.0cm}\epsfig{figure=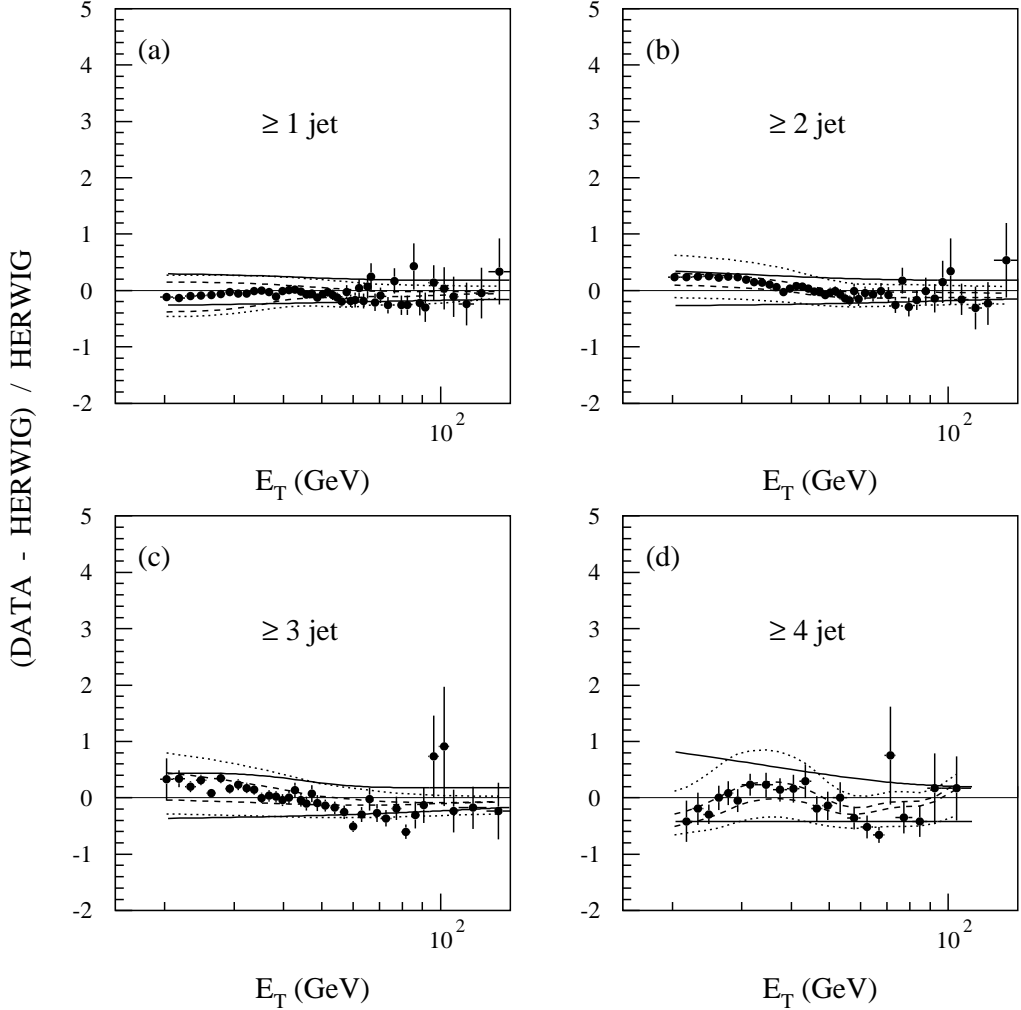
,width=15cm}}
\vspace {0.5cm}
\caption{
(Data $-$ {\sc herwig})/{\sc herwig} as a function of
the transverse energy of the leading jet for (a) single-jet inclusive,
(b) two-jet inclusive, (c) three-jet inclusive, and (d) four-jet
inclusive event samples. The relative systematic uncertainties in the 
cross section corresponding to the energy calibration added in quadrature 
with 15\% uncertainty in luminosity are shown by the solid lines. The 
uncertainty in the ratio (Data $-$ MC) / MC from energy and angle 
smearing is shown by the dashed lines. The total uncertainty on the ratio 
is shown by the dotted lines.
}
\label{et_prim_sys_angsm_hrw}
\end{center}
\end{figure}

The normalised theory is in agreement with the data for all of the  
jet samples over the entire $E_T$ interval. A detailed comparison is 
shown in Figs.~\ref{et_prim_sys_angsm} and ~\ref{et_prim_sys_angsm_hrw}. 
Here the simulations have been brought into agreement with the data by 
selecting parameters that enhance low $E_T$ jet production. In the case 
of {\sc pythia}, the core of the hadronic matter 
distribution\cite{ref_pythia} has been increased to the 
fraction 0.32. An increased core fraction (the parameter PARP(83))
leads to enhancement of the multiple interaction rate\cite{ref_pythia}, 
which tends to produce events with large multiplicity because of 
additional radiated low energy jets and underlying event energy. 
In the case of {\sc herwig}, the minimum transverse momentum for the
hard subprocesses has been set to 3.7 GeV. A decreased minimum 
transverse momentum (the parameter PTMIN) leads to increased soft 
underlying event contributions. The default values for these parameters 
are PARP(83)=0.5 and PTMIN=10$\rm ~GeV$. Variation of these values by 
more than 15\% leads to disagreement with the low $E_T$ data. 
Other parameters, when varied from their default values, do not change 
the distributions significantly. 

Figures~\ref{et_prim_sys_angsm} and ~\ref{et_prim_sys_angsm_hrw}
show the fractional difference (Data $-$ MC) / MC for the
$E_T$ spectra in Fig.~\ref{et_angsm} with the uncertainties
arising from jet-energy calibration and resolutions.  
The systematic uncertainty on the cross section is due primarily to
the uncertainty in the energy calibration. This uncertainty can be 
estimated by considering cross sections derived with $\pm 1$ 
standard-deviation corrections to the jet energy scale. 
The same procedure can be used to derive the uncertainties due to jet 
energy and angular resolutions in the MC. At $E_T$ $=$ 25 GeV, 
the uncertainty in the three-jet cross section due to calibration of 
the data is 39\%, and uncertainties in the MC due to energy and angular 
resolutions are 19\% and 7\%, respectively. 
\newpage
\vspace*{5.0cm}
\newpage
\vspace*{5.0cm}
\newpage
\vspace*{5.0cm}
\newpage
\noindent 
The uncertainty from energy 
resolution represents the dominant uncertainty in the MC. 
In Figs.~\ref{et_prim_sys_angsm} and
~\ref{et_prim_sys_angsm_hrw}, the relative systematic uncertainties in 
the cross section corresponding to the energy calibration added in 
quadrature with 15\% uncertainty in luminosity are shown by the solid 
lines. The uncertainty in the ratio (Data $-$ MC) / MC from energy and 
angle smearing is shown by the dashed lines. The total uncertainty on 
the ratio is shown by the dotted lines. As shown in 
Figs.~\ref{et_prim_sys_angsm} and
~\ref{et_prim_sys_angsm_hrw}, both {\sc pythia} and {\sc herwig}
describe the data quite well.

\section{Transverse energy and azimuthal distributions}
\label{sec:Sum}

\begin{figure}[h]
\vspace {-4.0cm}
\begin{center}
\mbox{\hspace*{0.0cm}\epsfig{figure=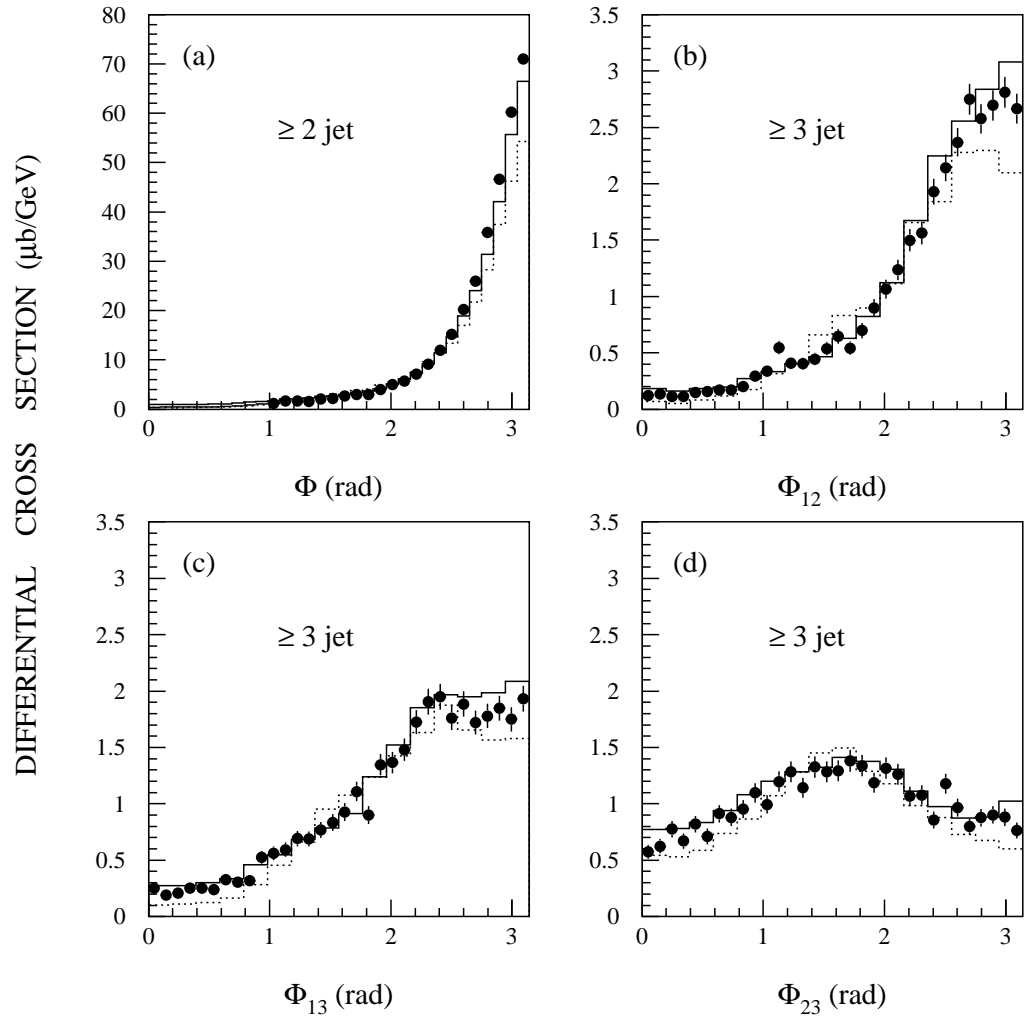,width=15cm}}
\vspace {0.5cm}
\caption{
Distributions of the relative azimuthal angle between two jets in 
(a) two-jet inclusive events and in three-jet inclusive events (b--d). 
Jets are ordered by their transverse energies. The {\sc pythia}
predictions are indicated by the solid histograms and the {\sc herwig}
predictions by the dotted histograms.
}
\label{azim_angsm}
\end{center}
\end{figure}

To explore features of three- and four-jet production, we turn to
observations of relative azimuthal distributions, distributions 
in summed transverse momenta, and three-jet studies. In
Fig.~\ref{azim_angsm}(a) we plot the azimuthal difference between
the leading two jets in events with two or more jets.
Figures~\ref{azim_angsm}(b--d) show the azimuthal difference
between the first and second, first and third, and second and
third highest-$E_T$ jets in three-jet events. In
Fig.~\ref{azim_angsm}(a) we see the strong anticorrelation (in the
transverse plane) expected of two-jet events. The peak of the
distribution widens substantially in the three-jet sample
(Fig.~\ref{azim_angsm}(b--d)). The peaks correspond to the
kinematic constraint of transverse momentum conservation for jets
produced in hard QCD subprocesses. {\sc pythia} (normalized as in
Fig.~\ref{et_angsm}) approximates the observed three-jet cross
section and shapes. However, small discrepancies with {\sc herwig}
(also normalized as in Fig.~\ref{et_angsm}) are evident.

\begin{figure}[h]
\vspace {-4.0cm}
\begin{center}
\mbox{\hspace*{0.0cm}\epsfig{figure=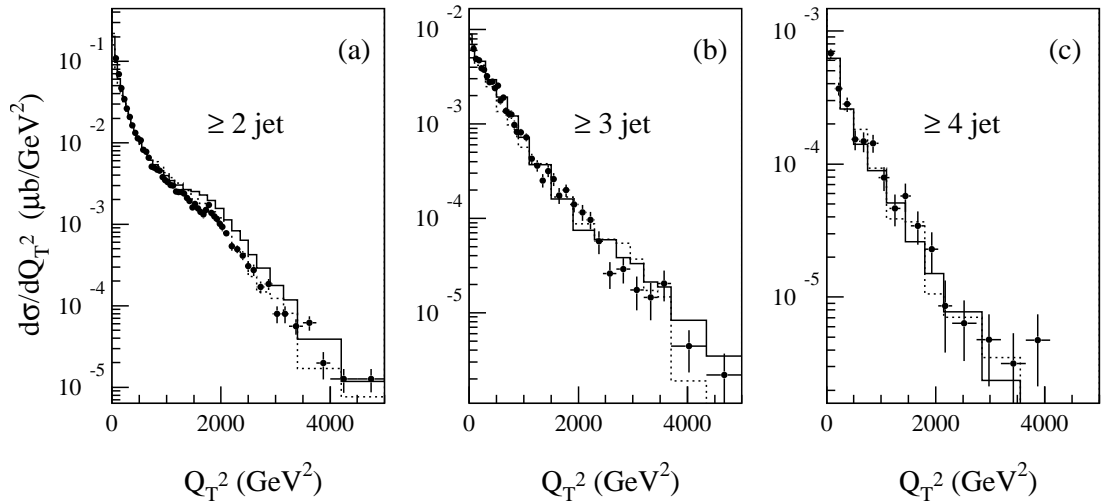,width=15cm}}
\vspace {0.5cm}
\caption{
Distributions of the square of the summed vector
transverse momenta $Q_{T}^2$, for (a) two-jet inclusive, (b) three-jet
inclusive, and (c) four-jet inclusive event samples. The {\sc
pythia} predictions are indicated by the solid histograms and the
{\sc herwig} predictions by the dotted histograms.
}
\label{q_T_angsm}
\end{center}
\end{figure}

Distributions of the square of the summed vector transverse
momenta of jets $Q_{T}^2=({\bf E}_{T1}+{\bf E}_{T2} +\cdots +{\bf
E}_{Tn})^2$ in Fig.~\ref{q_T_angsm} show significant
imbalance of the transverse momenta for $n$ leading jets. If events
at large $Q_{T}^2$ are removed by requiring balanced
transverse energy, the corresponding three- and four-jet cross
sections of Fig.~\ref{et_angsm} decrease at small $E_T$. The
shoulder at $Q_{T}^2 \approx 1600~\rm GeV^2$ in
Fig.~\ref{q_T_angsm}(a) can be eliminated by restricting the event 
sample to just two jets with $E_T$ above 20 GeV, and no
other jets between 8 and 20 GeV. This shoulder can consequently be
associated with higher-order radiation.

\begin{figure}[h]
\vspace {-4.0cm}
\begin{center}
\mbox{\hspace*{0.0cm}\epsfig{figure=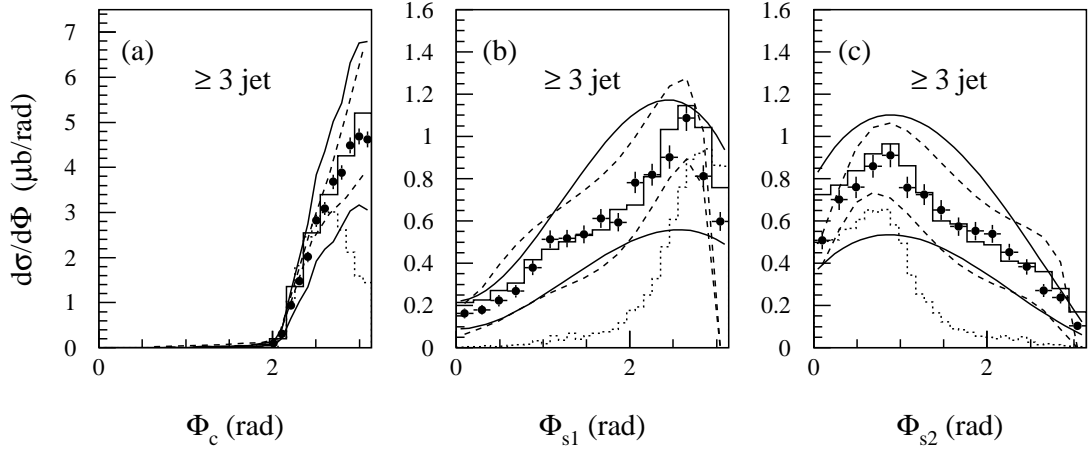
,width=15cm}} 
\vspace {0.5cm} 
\caption{ 
Azimuthal distributions
between the leading jets in 3-jet events. The data is shown by the
closed circles. Panel (a) shows the azimuthal separation between
the two jets with the minimum summed transverse energy. Panel (b)
shows the azimuthal separation between the third leading jet and
the first jet of the minimum transverse energy pair. Panel (c)
shows the azimuthal separation between the third leading jet and
the second jet of the pair. {\sc pythia} is given by the solid
histograms, {\sc jetrad} is shown by the dotted histograms. The 
uncertainties associated with energy calibration and luminosity are 
shown by the solid lines. Uncertainties from the energy resolution 
are shown by dashed lines. 
} 
\label{phic_phi_angsm}
\end{center}
\end{figure}

\begin{figure}[h]
\vspace {-4.0cm}
\begin{center}
\mbox{\hspace*{0.0cm}\epsfig{figure=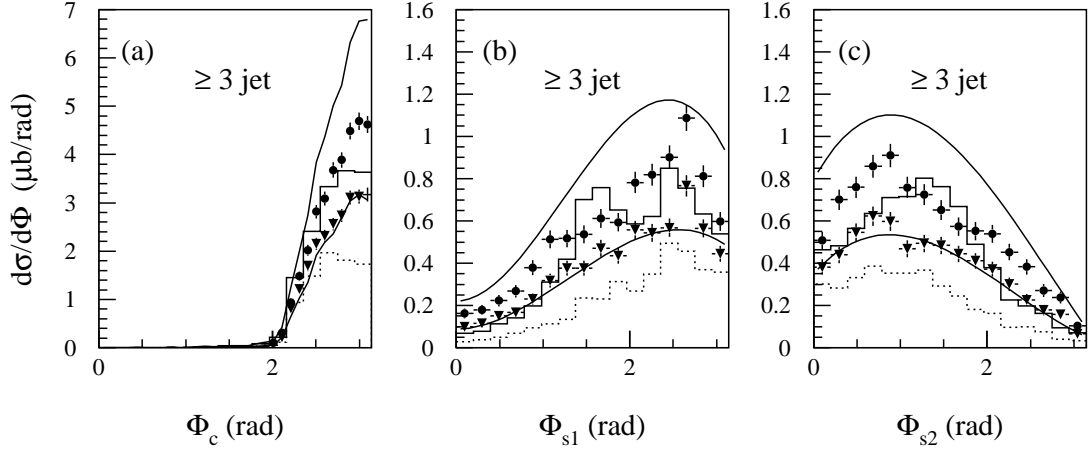
,width=15cm}} 
\vspace {0.5cm} 
\caption{ 
Azimuthal distributions
between the leading jets in 3-jet events. The data is given by the
closed circles (all jets) and by the closed triangles (the jets
overlapped with more than one jet are excluded). Panel (a) shows
the azimuthal separation between the two jets with the minimum
summed transverse energy. Panel (b) shows the azimuthal separation
between the third leading jet and the first jet of the minimum
transverse energy pair. Panel (c) shows the azimuthal separation
between the third leading jet and the second jet of the pair. {\sc
herwig} is given by the solid histograms (all jets), and the
dotted histograms (the jets overlapped with more than one jet are
excluded). The uncertainties associated with energy calibration and 
luminosity are shown by the solid lines. 
} 
\label{phic_phi_angsm_hrw}
\end{center}
\end{figure}

To find the pair of jets $\{i,j\}$ most likely to originate from
the hard interaction (rather than from gluon bremsshtrahlung), we
define the scaled summed dijet vector transverse momentum: ${\bf
q}_{ij}=({\bf E}_{Ti}+{\bf E}_{Tj})/(E_{Ti}+E_{Tj})$. We choose
the pair with the smallest magnitude of this vector and in
Figs.~\ref{phic_phi_angsm}(a) and \ref{phic_phi_angsm_hrw}(a) plot
the distribution of the relative azimuthal angle $\Phi _c$ between
the jets in that pair. The data, {\sc pythia}, and {\sc herwig}
show a narrow maximum in the region where two jets from the hard
scatter appear back-to-back ($\Phi_c = \pi$). The prediction from
{\sc jetrad} is peaked away from $\Phi_c \approx \pi$ because only
one extra jet is present.

Figures~\ref{phic_phi_angsm}(b,c) and
Fig.~\ref{phic_phi_angsm_hrw}(b,c) show the azimuthal separation
of the third jet from each of the two jets that correspond to the
minimum $q_{ij}^2$. These distributions contain events only for
$\pi -\Phi _c \le 0.4$; that is, events in which the balanced jets
are essentially back-to-back. If the third jet were correlated
with the balanced jets, it would be observed nearby or opposite
the balanced jets. However, the data show the third jet to be
weakly correlated with the balanced jets, and emitted at all
angles. The uncertainties associated with energy calibration and
luminosity are shown by the solid lines in
Figs.~\ref{phic_phi_angsm} and ~\ref{phic_phi_angsm_hrw}.
Uncertainties from the energy resolution are shown by dashed lines
in Fig.~\ref{phic_phi_angsm}.
\newpage
\vspace*{5.0cm}
\newpage
\vspace*{5.0cm}
\newpage
\vspace*{5.0cm}
\newpage
\vspace*{5.0cm}
\newpage

We see that the data, {\sc pythia}, and {\sc herwig} have wider
distributions than {\sc jetrad}.  {\sc pythia} describes the data
quite well, while {\sc jetrad} fails. The agreement with {\sc pythia} 
has been achieved only with enhanced multiple parton interaction 
rates. {\sc herwig} demonstrates small qualitative disagreement with 
the shape of the azimuthal plot of Fig.~\ref{phic_phi_angsm_hrw}(b); 
the peak at $\pi/2$ is produced by jets reconstructed from the 
underlying-event energy\cite{ref_4} and grows quickly with small 
changes in PTMIN. Such jets are strongly overlapped with more that one 
jet. If jets overlapping two or more nearby jets are excluded, the 
{\sc herwig} shape in Fig.~\ref{phic_phi_angsm_hrw}(b) improves but 
the agreement shown in Fig.~\ref{phic_phi_angsm_hrw}(a) worsens. 
(The cone algorithm reconstructs jets from seed towers and may 
therefore reconstruct jets sharing energy. The reconstruction algorithm 
then merges or splits the energy encompassed in these overlapping 
jets\cite{ref_11_0}.) Elimination of these jets tends to suppress 
contributions from the soft underlying event. Soft interactions result 
in a wide distribution of particles throughout angular phase space. Jets 
reconstructed from these particles tend to be wider and of lower energy 
than more collimated partonic jets. Such jets often share a significant 
fraction of energy with similar, neighbouring jets and are merged into 
a single jet.

The shapes of the simulated distributions are sensitive to
modeling of the multiple parton interactions. Tuning of the
multiple interaction contribution in {\sc pythia} and the minimum
generated transverse momentum in {\sc herwig} are required for
good agreement. In particular, simulations with smaller
contributions from soft parton interactions show discrepancies
with the data.

\section {Conclusions}
\label{sec:Conc}

In this paper we showed comparisons between Monte Carlo and data for 
several characteristics of multiple jet events with a low jet-$E_T$ 
threshold. These comparisons included the leading jet transverse 
energy, the relative azimuthal angle between jets, and the summed 
vector transverse momenta of jets. Our data on multiple jet 
production at low $E_T$ agree with {\sc pythia} and {\sc herwig}. 
This is observed in the distributions of the
transverse energy of the leading jets (Fig.~\ref{et_angsm}),
azimuthal distributions (Fig.~\ref{azim_angsm}), in the square of
the summed vector transverse momenta $Q_{T}^2$
(Fig.~\ref{q_T_angsm}), and in the three-jet angular distributions
that suggest the presence of a weakly correlated jet
(Figs.~\ref{phic_phi_angsm}, ~\ref{phic_phi_angsm_hrw}). {\sc jetrad} 
cannot adequately describe the angular distributions of the three 
leading jets in three jet events.

\section {Acknowledgments}
\label{Ackn}

%
%

%
We thank the staffs at Fermilab and collaborating institutions, 
and acknowledge support from the 
Department of Energy and National Science Foundation (USA),  
Commissariat  \` a L'Energie Atomique and 
CNRS/Institut National de Physique Nucl\'eaire et 
de Physique des Particules (France), 
Ministry for Science and Technology and Ministry for Atomic 
   Energy (Russia),
CAPES and CNPq (Brazil),
Departments of Atomic Energy and Science and Education (India),
Colciencias (Colombia),
CONACyT (Mexico),
Ministry of Education and KOSEF (Korea),
CONICET and UBACyT (Argentina),
The Foundation for Fundamental Research on Matter (The Netherlands),
PPARC (United Kingdom),
Ministry of Education (Czech Republic),
A.P.~Sloan Foundation,
and the Research Corporation.
%

%

\newpage


\end{document}